\documentclass[letter,aps,reprint,twocolumn,showpacs,preprintnumbers,amsmath,amssymb]{revtex4}
\usepackage{amsmath,amssymb,graphicx}
\usepackage {amssymb}
\usepackage{multirow}
\usepackage{color}
\newcommand{\nc}{\newcommand}
\nc{\ba}{\begin{eqnarray}} \nc{\ea}{\end{eqnarray}}
\nc{\be}{\begin{equation}} \nc{\ee}{\end{equation}}

\newcommand{\eea}{\end{eqnarray}}
\newcommand{\bea}{\begin{eqnarray}}

\newcommand\s{\sigma}

\newcommand\e{\epsilon}

\newcommand\om{\omega}
\newcommand\D{{\cal D}}
\newcommand\de {\partial}

\nc{\ga}{\gamma} \nc{\x}{{\bf x }} \nc{\kk}{{\bf k }} \nc{\f}{{\bf f
}} \nc{\T}{ \theta (s_i (t)- \s) } \nc{\TT}{ \theta (s_i (t_{ r \, i
} )- \s) } \nc{\br}{   (s_i (t)- \s)  } \nc{\fa}{\phi_1}
\nc{\fb}{\phi_2}
\def\Lag{{\cal L}}
\input{epsf.sty}

\begin{document}
\title{Effects of a Thermal Bath of Photons on Embedded String Stability}

\author{Johanna Karouby and Robert Brandenberger}

\affiliation{$^1$Department of Physics, McGill University,
Montr\'eal,
QC, H3A 2T8, Canada}

\begin{abstract}

We compute the corrections of thermal photons on the
effective potential for the linear sigma model of QCD.
Since we are interested in temperatures lower than the
confinement temperature, we consider the scalar fields
to be out of equilibrium. Two of the scalar
field are uncharged while the other two are charged under the
U(1) gauge symmetry of electromagnetism. We find that the
induced thermal terms in the effective potential can stabilize
the embedded pion string, a string configuration which is
unstable in the vacuum. Our results are applicable in
a more general context and demonstrate that embedded
string configurations arising in a wider class of field
theories can be stabilized by thermal effects. Another
well-known example of an embedded string which can be
stabilized by thermal effects is the electroweak Z-string.
We discuss the general criteria for thermal stabilization
of embedded defects.

\end{abstract}

\maketitle
\section{Introduction}

 Topological defects can play an important role in early universe
cosmology (see e.g. \cite{ShellVil, HK, RHBrev} for overviews). On one hand, particle physics models which yield 
defects such as domain walls which have problematic and unobserved 
effects can be ruled out. On the other hand, topological
defects may help explain certain cosmological observations. They
could contribute to structure formation or generate primordial
magnetic fields which are coherent on cosmological scales \cite{CSmag}.

The Standard Model of particle physics does not give rise
to topological defects which are stable in the vacuum. On the
other hand, it is possible to construct string-like
configurations which would be topological defects if certain
of the fields were constrained to vanish. If they are not,
then the defects are unstable in the vacuum. Such defects
are called ``embedded defects" (see \cite{embed} for an
overview). Two prime examples of such embedded defects
are the pion string arising in the low energy linear sigma
model of QCD \cite{pionstring}, and the electroweak Z-string
\cite{Zstring} arising in the electroweak theory.

Both the pion string and the electroweak Z-string arise
in models with two complex scalar fields, one of them
uncharged with respect to the $U(1)$ gauge field of
electromagnetism, the second charged. In terms of real
fields, we have four real scalar fields $\phi_i, i = 0 , .. , 3$
with a bare potential of the form
\be
V(\phi) \, = \, \frac{\lambda}{4} \bigl( \phi^2 - \eta^2 \bigr)^2 \, ,
\ee
where 
$\phi^2 \, = \, \sum_{i = 0}^3 \phi_i^2 \, .$
In both physical examples, two of the fields ($\phi_0$ and $\phi_3$)
are uncharged whereas the two others are charged.
Since the vacuum manifold of this theory is ${\cal M} = S^3$,
there are no stable topological defects, only $\Pi_3$ defects
which in cosmology are called textures \cite{Neil}.

However, the presence of an external electromagnetic field
breaks the symmetry since only two of the fields couple
to the photon field. Interactions with the photon field
lift the potential in the charged field directions, leading
to a reduced symmetry group $G$ which is
\be
G \, = \, U(1)_{global} \times U(1)_{gauge}
\ee
instead of $O(4)$. The vacuum manifold becomes 
a circle ${\cal M} = S^1$ corresponding to
\be \label{redvacman}
\phi_1 = \phi_2 = 0, ~~~~ \phi_0^2 + \phi_3^2 = \eta^2 \, .
\ee
It is possible to construct embedded cosmic string solutions
which are topological cosmic strings of the reduced
theory with $\phi_1 = \phi_2 = 0$. The field configuration of
such a string (centered
at the origin of planar coordinates and extended along the
z-axis) takes the form
\be \label{string}
\Phi(\rho, \theta) \, = \, f(\rho) \eta e^{i \theta} \, ,
\ee
where the complex electrically neutral field is $\Phi = \phi_0 + i \phi_3$.
 $f(\rho)$ is a function which interpolates between $f(0) = 0$ and
$f(\rho) = 1$ for $\rho \rightarrow \infty$ with a width which is of the
order $\lambda^{-1/2} \eta^{-1}$. In the above, $\rho$ and $\theta$ are
the polar coordinates in the plane perpendicular to the z-axis.

In \cite{Nagasawa}, a plasma stabilization mechanism for the pion
string and the electroweak Z-string was proposed. The argument was
based on interpreting the terms in the covariant derivative which couple
the charged scalar field $\pi^+ = (1 / \sqrt{2}) (\phi_1 + i \phi_2)$ 
to the gauge field as a term which, if the gauge field is in 
thermal equilibrium, will add a term proportional to 
$\delta V \, \sim \, e^2 T^2 |\pi^{+}|^2,$
to the effective potential of the scalar field sector. This lifts the potential
in the direction of the charged scalar fields, leaving us with a reduced
vacuum manifold given by (\ref{redvacman}).

In this paper we put this suggested stabilization mechanism on a firmer
foundation, focusing on the example of the pion string. The setting
of our analysis is the following: we are interested in temperatures
below the chiral symmetry breaking transition. Hence, the scalar fields
are out of thermal equilibrium. However, the photon field is in thermal
equilibrium. This is not the usual setting for finite temperature
quantum field theory (since not all fields are in thermal equilibrium)
and hence non-standard techniques are required.

We compute the effective potential for the scalar fields
obtained by integrating out the gauge field, taking it to be in
thermal equilibrium. To do so we use a functional integral in which the
time domain is Euclidean and ranges from $0$ to $\beta$, where $\beta$
is the inverse temperature. We find that the resulting scalar field effective
potential has a broken symmetry and a vacuum manifold given by
(\ref{redvacman}). There is hence an energetic barrier which has to
be overcome to destroy a pion string.

\section{The Pion String}
 The cosmological context of this work is Standard Big Bang Cosmology.
At about 10 microseconds after the Big Bang a phase
transition from the quark-gluon plasma to a hadron gas
is expected to have taken place at a temperature of about
$T_{QCD} \sim 150 - 200 {\rm MeV}$. Below this critical
temperature, the physics of hadrons can be well described
by a linear sigma model of four scalar fields which
we collectively denote by $\phi$,
three of them representing the pions. If we make
the standard assumption that the relevant bare quark
masses vanish, i.e. $m_u = m_d= 0$, where the subscripts
stand for the up and down quark, respectively, we know
that the effective action (in vacuum) will have a
$SU(2)$ symmetry which is spontaneously broken by
a potential $V(\phi)$. This spontaneous symmetry
breaking leads to a mass for the fourth scalar field,
the so-called sigma field.
 
In the setup described above, the ``vacuum manifold'' ${\cal M}$, i.e.
the set of field configurations which minimize the potential,
is nontrivial and takes the form of a 3-sphere $S^3$.
There are no stable topological defects associated with this
symmetry breaking, only global textures which are not
stable. In particular, since the first homotopy group of
the vacuum manifold is trivial, i.e. $\Pi_1({\cal M}) = 1$,
there are no stable cosmic strings.

However, as discussed in \cite{pionstring}, it is possible
to construct embedded strings for which the charged pion
fields are set to zero, and the two neutral fields (the neutral
pion and the sigma fields) form a cosmic string
configuration. This is the so-called ``pion string''.

Since two of the four scalar fields are charged (the two
charged pion fields) while the remaining two are neutral,
turning on the electromagnetic field will destroy the $O(4)$
symmetry and will break it down to $U(1)_{global} \times U(1)_{local}$. The second
factor corresponds to rotations of the charged complex
scalar field, the first to a rotation of the neutral one.
The pion string can be viewed as the cosmic string configuration
associated with the first $U(1)$.

As a toy model for the analytical study of the stabilization of
embedded defects by plasma effects we consider the chiral limit of the
QCD linear sigma model, involving the sigma field $\sigma$ and the
pion triplet ${\vec \pi} = (\pi^0, \pi^1, \pi^2)$, given by the
Lagrangian
\begin{equation}
\label{lag1}
{\cal L}_0 \, = \, {1 \over 2} \partial_{\mu} \sigma \partial^{\mu}
\sigma + {1 \over 2} \partial_{\mu} {\vec \pi} \partial^{\mu} {\vec
\pi} - {\lambda \over 4} (\sigma^2 + {\vec \pi}^2 - \eta^2)^2 \, ,
\end{equation}
where $\eta^2$ is the ground state expectation value of 
$\sigma^2 + {\vec \pi}^2$. In the following, we denote the 
potential in (\ref{lag1}) by $V_0$.

Two of the scalar fields, the $\sigma$ and $\pi_0$, are electrically
neutral, the other two are charged. Introducing the coupling to
electromagnetism, it is convenient to write the bosonic sector
${\cal L}$ of the resulting Lagrangian in terms of the complex scalar
fields
\begin{equation}
\pi^+ \, = {1 \over {\sqrt{2}}} (\pi^1 + i \pi^2) ,  \,\,
\pi^- \, = {1 \over {\sqrt{2}}} (\pi^1 - i \pi^2) \, .
\end{equation}
According to the minimal coupling prescription we obtain $$
{\cal L} \, = \, {1 \over 2} \partial_{\mu} \sigma \partial^{\mu}
\sigma + {1 \over 2} \partial_{\mu} \pi^0 \partial^{\mu} \pi^0 +
D_{\mu}^+ \pi^+ D^{\mu -} \pi^--\frac{1}{4} F_{\mu \nu} F^{\mu \nu}   + V_0 \, ,$$
where
$D_{\mu}^+ \, = \, \partial_{\mu} + i e A_{\mu}$ , $D_{\mu}^-
\, = \, \partial_{\mu} - i e A_{\mu} \, .$

Effective pion-photon interactions appear through the covariant 
derivative. They break the $O(4)$ symmetry
which the Lagrangian would have in the absence of the gauge
field.

In the following we work in terms of the two
complex scalar fields
\ba
&\Phi \, = \, \s + i \pi_0   \nonumber \\
&\pi_c \, = \, \pi_1 + i \pi_2 \, ,
\ea
the first of which is electrically neutral, the second charged.

The minimum of the potential can be obtained for $\langle\pi_c\rangle=0$ and $\langle \Phi\rangle = \eta$ and electromagnetism 
is unbroken. In that case, the vacuum manifold $S^3$ reduces to $S^1$ 
and some string configurations exist.
They are not topologically stable since they can unwind by
exciting the charged fields, i.e. $\langle\pi_c \rangle\neq 0$. This string
solution is the ``pion string''. The field $\Phi$ vanishes in
the center of the string (the charged field
vanishes everywhere)  and this implies that there is trapped
potential energy along the string.

Note that if we distort the field configuration to have
$\langle\pi_c\rangle \neq 0$, then the U(1) of electromagnetism 
gets broken and there is a magnetic flux of $\frac{2\pi}{e}$ 
in the core of the string. To see how this flux may arise,
consider as starting point the pion string configuration
with $\pi_c $ vanishing everywhere.
To see the effect of the charged field, let us now consider
exciting a constant $\pi_c $ in the core of the string
and see how the potential energy evolves in this case.
We easily find that as long as $|\pi_c| \ll \eta $, we get an 
increase in the potential energy in the core of the string:
\be
V(\phi=0, \pi_c) \, \simeq \, V_0(0, 0) + \sqrt {\lambda V_0} |\pi_c|^2
\ee
and the lowest potential energy configuration is obtained for $\pi_c=0$.
Based on potential energy arguments alone we would infer
that we could get a stable string. However, the kinetic and gradient 
energies lead to an instability of the vacuum pion string
configuration.

We want to consider the effect of a photon plasma on the stability 
of the pion string. Here we consider an ultrarelativistic plasma 
of photons. The effective Lagrangian that
describes such a plasma is \cite{Braaten:1992jj} :
\be
\Lag_{eff} \;=\; \Lag_{QED} \;+\; \Lag_\gamma
\label{LeffQED} 
\ee
We do not consider the electrons in the Lagrangian for QED, but
simply take
\be
\Lag_{QED} \; = \; - {1 \over 4} F_{\mu \nu} F^{\mu \nu} \, ,
\ee
where $F_{\mu \nu}$ is the field strength associated with the
electomagnetic 4-potential $A_{\mu}$.

The plasma terms in the effective Lagrangian (\ref{LeffQED}) are
\be
\Lag_\gamma \;=\; {3 \over 4} \; m_\gamma^2 \; F_{\mu \alpha}
  \left< { K^\alpha K^\beta \over (K \cdot \partial)^2 } \right>
  {F^\mu}_\beta \;,
\label{Lgamur}  
\ee
where $K^{\alpha}$ is the four-vector which represents the
momentum of the hard field in the loop, and $m_\gamma$
is the thermal photon mass which is given by 
\be
m_\gamma^2 \;=\; \frac{e^2}{9}
  \left( T^2 + {3 \over \pi^2} \mu^2 \right) \;,
\label{mgam}
\ee
and $\mu$ is the quark chemical potential.

What really matters is not the plasma behaviour but rather its 
influence on a charged scalar field.
We will use the ``Hard Thermal Loop'' formalism, 
that is we work in the limit where the plasma temperature is much higher
than any momentum or mass scale in the problem.
In our case we have the following for a scalar field $\phi$ in the 
fundamental representation of the gauge group \cite{Braaten:1991gm}:
 \be
\Lag_s\;=\; {3 \over 4} \; m_s^2 \; \phi ^\dagger
  \left< { D^2 \over (K \cdot D)^2 } \right> \phi
 \label{Ls}  
\ee
This reduces to 
\be
\Lag_s=m_s^2 \phi^\dagger \phi
\ee
where $K$ is the moment of the hard field in the loop and  
$m_s=\frac{e T}{2}$ is the thermal scalar mass 
(see \cite{Kraemmer:1995qe} for details).

Applying the above to our charged field $\pi_c$, 
the effective Lagrangian becomes
\ba
\label{Leff}
{\cal L} \, &=& - {1 \over 4} F_{\mu \nu} F^{\mu \nu}+  
 {1 \over 2} \partial_{\mu} \sigma \partial^{\mu}
\sigma + {1 \over 2} \partial_{\mu} \pi^0 \partial^{\mu} \pi^0 \nonumber \\
& & + D_{\mu}^+ \pi^+ D^{\mu -} \pi^- - V_0 -\frac{e^2 T^2 }{4}\pi^+\pi^- \, .
\ea
This gives rise to a new effective potential while retaining the 
gauge invariance of the Lagrangian:
\be
\label{Veff}
V_{eff} \, = \, {\lambda \over 4} (\sigma^2 
+ {\vec \pi}^2 - \eta^2)^2+\frac{e^2 T^2 }{4}\pi^+\pi_{-} \, ,
\ee
 where $ 
\frac{e^2 T^2 }{4}\pi^+\pi^{-} \, = \, 
\frac{e^2 T^2 }{8}(\pi_1^2+\pi_2^2)=\frac{e^2 T^2 }{8}|\pi_c|^2 \,. $
Based on potential energy considerations, the induced terms
in the effective potential should lead to the stabilization
of pion strings. 

\section{Effective potential computation}

An improved way to study the stability of the pion string
in the presence of a thermal bath is to determine the
effective potential of the scalar fields (which are out
of thermal equilibrium) in the presence of a thermal
bath of photons. This effective potential can be obtained
by computing the finite temperature functional integral
over the gauge field, treating the scalar fields as
external out-of-equilibrium classical ones. They
are out of thermal equilibrium since their
masses are heavy compared to the temperature if we
are below the critical temperature. Note also that 
string configurations are out-of-equilibrium states
below the Ginsburg temperature (which is the temperature
slightly lower than the critical temperature when the
defect network freezes out during the symmetry breaking
phase transition \cite{ShellVil, HK, RHBrev}). 

We make use of the imaginary time formalism \cite{Dolan:1973qd} of
thermal field theory (see e.g. \cite{Brandenberger:1984cz} for a review). 
 We work in Euclidean space-time: $t \rightarrow i\tau$ and 
$\tau : 0 \rightarrow T^{-1}=\beta$.

The starting point is the action
\bea
 &S[A^{\mu}, \Phi, \pi_c] = \int d^4x [{1 \over 2} \partial_{\mu} \Phi\partial^{\mu} \Phi+ {\lambda \over 4}(|\Phi|^2 + |\pi_c|^2-\eta^2)^2 ]\nonumber \\
&+\int_0^\beta d\tau \int d^3x [ D_{\mu}^+ \pi^+ D^{\mu -} \pi^--\frac{1}{4} F_{\mu \nu} F^{\mu \nu} ]
\eea
where the space integral runs from $-\infty$ to $+\infty$.

In the imaginary time formalism of thermal field
theory, the integration over four-momenta is carried out
in Euclidean space with $k_0 = ik_4$, this means that
the transition from zero temperature field theory is
obtained via $  \int \frac{d^4k}{(2\pi)^4} \rightarrow i \int \frac{d^4k_E}{(2\pi)^4} $.
Next, we recall that boson energies take discrete values, namely
$k_4=\omega_n=2n \pi T$ with $n$ an integer, and thus 
\bea
 \int \frac{d^4k_E}{(2\pi)^4} \rightarrow T \sum_n \int \frac{d^3k}{(2\pi)^3}.
\eea

We use this Matsubara mode decomposition for the gauge field only, because it is the only field in thermal
equilibrium. 

 The standard definition for the effective potential is based on the Legendre transform of the
generating functional (see \cite{Brandenberger:1984cz, Laine:2002zu} for 
reviews). However, the finite temperature effective potential with the scalar fields viewed
as classical background fields can also be defined as
\bea \label{Veff2}
Z[T]&=&\int \D \Phi \D\pi_c \D A^{\mu} e^{-S[A^{\mu}, \Phi,\pi_c]} \nonumber \\ &=&
\int \D \Phi \D\pi_c e^{-S[\Phi, \pi_c]} 
e^{-\frac{V_{eff}( \Phi,\pi_c) V}{T}}  
\eea
where $S[\Phi, \pi_c]$ is the gauge field independent part
of the (non-euclidean) action, $V$ is the volume of the system and 
$\int d\tau d^3\bf{x}=\frac{V}{T}$.

To evaluate the total partition function $Z[T]$ of the system 
we work in the covariant Feynman gauge following the procedure
reviewed e.g. in \cite{Kapusta:2006pm}, according to which the 
two unphysical degrees of freedom of $A^{\mu}$ that correpond 
to the longitudinal and timelike photons are cancelled by the 
ghost and anti-ghost fields $c $ and $\bar{c}$: 
\bea
&& Z[T]=
 \int \D \Phi \D \pi_c \D c\D \bar{c} \D A_{\mu}
 e^{ - S[\Phi,\pi_c]} \\ \nonumber
  &&\times e^{ -\int_0^\beta d\tau \int d^3x \ \bar{c} (-\de^2-e^2|\pi_c|^2)c } 
e^{ -\int_0^\beta d\tau \int d^3x \ \frac{1}{2}  A_{\mu} (\de^2+e^2 |\pi_c|^2) A_{\mu}}
 \label{Z1}
\eea
Here the summation of $A_{\mu}A_{\mu}$ is in Euclidean space since  
$A_{0}\rightarrow i A_{0}$. 
We can see from above that the gauge field obtains an effective mass 
equal to $m_{eff} \, = \, e|\pi_c| \,  $.
Now we  simply evaluate the Gaussian integration over the gauge field and the ghost fields.
\bea 
&&Z[T]= \int \D \Phi \D\pi_c e^{- S[\Phi,\pi_c]}  \nonumber \\
&&\times e^{2\frac{1}{2}Tr[\ln(\omega_n^2+{\mathbf{k}}^2+m_{eff}^2)]} e^{-4\frac{1}{2}Tr[\ln(\omega_n^2+{\mathbf{k}}^2+m_{eff}^2)]} \nonumber \\
&&Z[T]= \int \D \Phi \D\pi_c e^{- S[\Phi,\pi_c]} e^{-Tr[\ln(\omega_n^2+{\mathbf{k}}^2+m_{eff}^2)]} \,
 \label{Z2}
\eea
Comparing (\ref{Z2}) with the definition (\ref{Veff2})
of the effective potential we find
\bea
V_{eff}(\Phi,\pi_c,T)=V_0+\lim_{V\rightarrow \infty} \frac{T}{V}\sum_{n \in \mathbb{Z}}{\ln(\omega_n^2+{\mathbf{k}}^2+m_{eff}^2)}+{\rm cst} \nonumber\\
={\lambda \over4}(|\Phi|^2 +|\pi_c|^2-\eta^2)^2 +
2\int\frac{d^3k}{(2\pi)^3} [\frac{\omega}{2}+ T  \ln(1-e^{-\frac{\om}{T}})] \, \ \ 
\label{V3}
\eea
where 
$\om=  \sqrt{ {\bf{k}}^2+m_{eff}^2 }$ and $
V_0 \, = \, {\lambda \over 4}(|\Phi|^2+|\pi_c|^2-\eta^2)^2$.
The thermal part, $J(m_{eff},T)=\int\frac{d^3k}{(2\pi)^3}T\ln(1-e^{-\frac{\om}{T}})$, \ \ \ \ admits a high-temperature expansion :
\bea \label{Texp}
J(m_{eff},T) \, &=& \frac{T}{2}\sum_n\int\frac{d^3k}{(2\pi)^3} 
\ln(\omega_n^2+{\mathbf{k}}^2+m_{eff}^2) \nonumber\\
&\simeq& \, -\frac{\pi^2T^4}{90} + \frac{m_{eff}^2T^2}{24} -
\frac{m_{eff}^3T}{12\pi} \\
& & \, -\frac{m_{eff}^4}{32\pi^2} \left[\ln\left(\frac{m_{eff} e^{\gamma_{E}}}{4\pi T}\right)-\frac{3}{4}\right]
+ {\cal O}(\frac{m_{eff}^6}{T^2}) \, . \nonumber
\eea
The zero temperature part, $J_{0}(m)$, is UV divergent :
\bea
J_{0}\left(m\right) &=& 
\left.\int\frac{d^{3}p}{\left(2\pi\right)^{3}}\frac{\omega}{2}
\right|_{\omega=\sqrt{p^{2}+m^{2}}} \, \nonumber\\
&=& \, -\frac{m^{4}\mu^{-2\epsilon}}{64\pi^{2}}
\left[\frac{1}{\epsilon}+\ln\frac{\overline{\mu}^{2}}{m^{2}}+\frac{3}{2}+O\left(\epsilon\right)\right] \, .
\eea
The renormalized value of this integral has been obtained using the
$\overline{MS}$ renormalization parameter $\bar{\mu}$.
 $d=3-2\e$ is the dimension of the momentum integral.
Considering renormalization of coupling constants as well, will give $O(\hbar)$ 
corrections to the potential, but should not change the topology of the vacuum manifold. 
For example, if the vacuum manifold is a circle it could become
a titled circle but since these are small effects, this will not affect the presence of a topological 
defects.

At high-temperatures we can truncate the series in (\ref{Texp}) and get
\bea
V_{eff} (\Phi,\pi_c,T)\, = \, {\lambda \over 4} (|\Phi|^2+|\pi_c|^2-\eta^2)^2 
 - \frac{\pi^2 T^4}{45} + \frac{e^2|\pi_c|^2 T^2}{12}  \nonumber\\
 -\frac{e^3|\pi_c|^3  T}{6\pi}
- \frac{e^4 |\pi_c|^4}{16\pi^2} \left[\ln\left(\frac{e |\pi_c| e^{\gamma_{E}}}{4\pi T}\right)-\frac{3}{4}\right] \nonumber
\eea
where we neglect terms of order $\frac{e^6|\pi_c|^6}{T^2}$. This potential is also
the approximate potential close to the (0,0) point of the 
$(\Phi,\pi_c)$-plane.
For consistency with the hard thermal loop approximation, we
can check that, when the gauge field takes on an effective mass, $m_{eff}$,
it has 3 polarizations instead of 2. This leads to the appearance of an overall
factor of $\frac{3}{2}$ for the thermal part of the effective potential.
The term  quadratic in temperature then becomes $m_{eff}^2 T^2/8$
as it appears in (\ref{Veff}).

The above computation shows that the effects of the photon
plasma create an energy barrier which lifts the scalar
field effective potential in direction of the charged fields.
In order to minize this effective potential, the charged fields must go to zero.
The ``effective'' vacuum manifold ${\cal M}$ is now no longer 
$S^3$ but rather
${\cal M} \, = \, S^1$ 
which has nontrivial first homotopy group and hence admits
stable cosmic string solutions which are precisely the
pion strings discussed earlier.

\section{Conclusions}

We have studied the effective potential for the scalar fields
of the low-energy effective sigma model of QCD in the presence
of a thermal bath of photons. We have shown that the
plasma effects break the symmetry of the theory, lift the
potential in direction of the charged pion fields, and
lead to an effective vacuum manifold which admits cosmic
string solutions, the pion strings. 

Our analysis puts the stabilization mechanism of \cite{Nagasawa}
on a firmer footing. It shows that if pion strings form,
they will be stabilized by plasma effects, at least at a classical
level. The stability of pion strings to quantum processes remains
to be studied.

The analysis in this paper applies not only to pion strings, but
equally to the corresponding embedded strings in the electroweak
theory, the Z-string. Thus, below the confinement scale the
Standard Model of particle physics admits two types of classically
stable embedded strings. 

Our arguments, however, are more general and apply to many theories
beyond the Standard Model. Given a theory with a multi-component
scalar field order parameter with one set of components which
are neutral, and a second set which are charged - neutral and
charged being with respect to the fields excited in the plasma.
Then topological defects of the theory with vanishing charged order
parameters become embedded defects of the full theory with the
property that they are stabilized in the early universe.

Stabilized embedded defects can have many applications in cosmology.
For example, stabilized pion strings provide  an explanation for
the origin and large-scale coherence of cosmological magnetic fields \cite{CSmag}.
On the other hand, stabilized embedded domain walls would lead to
an overclosure problem. Hence, theories admitting those types of
embedded defects would be ruled out by cosmological considerations.

\begin{acknowledgments}

We wish to thank Aleksi Kurkela and Guy Moore for useful discussions.
This work is supported in part by a NSERC Discovery Grant to RB and by
funds from the Canada Research Chair program. JK is supported by
an NSERC Postgraduate Scholarship.

\end{acknowledgments}

\end{document}